\documentclass[proceedings]{JHEP3}
\usepackage{amsfonts}
\usepackage{amsmath}
\usepackage{epsfig,multicol}

\newbox\mybox

\newcommand\fverb{\setbox\mybox=\hbox\bgroup\verb}
\newcommand\fverbdo{\egroup\medskip\noindent\fbox{\unhbox\mybox}\ }
\newcommand\fverbit{\egroup\item[\fbox{\unhbox\mybox}]}
\conference{...generic cubic $\mathcal{PT}$-symmetric non-Hermitian Hamiltonian}
\abstract{We investigate properties of the most general $\mathcal{PT}$-symmetric 
non-Hermitian Hamiltonian of cubic order in the annihilation and creation operators as
a ten parameter family. For various choices of the parameters we systematically
construct an exact expression for a metric operator and an isospectral Hermitian counterpart in the same 
similarity class by exploiting the isomorphism between operator and Moyal products.
We elaborate on the subtleties of this approach.
For special choices of the ten parameters the Hamiltonian reduces to various models
previously studied, such as to the complex
cubic potential, the so-called Swanson Hamiltonian or the transformed version of the
from below unbounded quartic $-x^4$-potential. In addition, it also reduces to various models
not considered in the present context, namely
the single site lattice Reggeon model and a transformed version of the massive sextic
$\pm x^6$-potential, which plays an important role as a toy model
to identify theories with vanishing cosmological constant.}

\title{Metrics and isospectral partners for the most generic cubic $\mathcal{%
PT}$-symmetric non-Hermitian Hamiltonian}
\author{Paulo E.G. Assis$^1$ and Andreas Fring$^{1,2}$ \\
$^1$Centre for Mathematical Science, City University London, \\
Northampton Square, London EC1V 0HB, UK\\
$^2$Department of Physics, University of Stellenbosch, 7602 Matieland, South Africa\\
E-mail: Paulo.Goncalves-De-Assis.1@city.ac.uk, A.Fring@city.ac.uk}

\input{tcilatex}

\begin{document}

\section{Introduction}

Non-Hermitian Hamiltonians are usually interpreted as effective Hamiltonians
associated with dissipative systems when they possess a complex eigenvalue
spectrum. However, from time to time also non-Hermitian Hamiltonians whose
spectra were believed to be \textit{real} have emerged sporadically in the
literature, e.g. the lattice version of Reggeon field theory \cite{RR3,RRR}. 
Restricting this model to a single site leads to a potential
very similar to the complex cubic potential $V=ix^{3}$. Somewhat later it
was found \cite{Cal} for the latter model that it possess a real spectrum on
the real line. More recently the surprising discovery was made \cite%
{Bender:1998ke} that in fact the entire infinite family of non-Hermitian
Hamiltonians involving the complex potentials $V^{n}=z^{2}(iz)^{n}$ for $%
n\geq 0$ possess a real spectrum, when its domain is appropriately continued
to the complex plane.

Thereafter it was understood \cite{Bender:1998ke,Bender:2002vv} that the
reality of the spectra can be explained by an unbroken $\mathcal{PT}$%
-symmetry, that is invariance of the Hamiltonian and its eigenfunctions
under a simultaneous parity transformation $\mathcal{P}$ and time reversal $%
\mathcal{T}$. In case only the Hamiltonian is $\mathcal{PT}$-symmetric the
eigenvalues occur in complex conjugate pairs. In fact, the $\mathcal{PT}$%
-operator is a specific example of an anti-linear operator for which such
spectral properties have been established in a generic manner a long time
ago by Wigner \cite{EW}. However, in practical terms one is usually not in a
position to know all eigenfunctions for a given non-Hermitian Hamiltoinian
and therefore one has to resort to other methods to establish the reality of
the spectrum. Since Hermitian Hamiltonians are guaranteed to have real
spectra, one obvious method is to search for Hermitian counterparts in the
same similarity class as the non-Hermitian one. This means one seeks
similarity transformations $\eta $ of the form 
\begin{equation}
h=\eta H\eta ^{-1}=h^{\dagger }=\eta ^{-1}H^{\dagger }\eta ~~\Leftrightarrow
~~H^{\dagger }=\eta ^{2}H\eta ^{-2},\qquad \text{for }\eta =\eta ^{\dagger }%
\text{.}  \label{1}
\end{equation}
Non-Hermitian Hamiltonians $H$ respecting the property (\ref{1}) are
referred to as pseudo-Hermitian \cite{Mostafazadeh:2002hb}. Besides these
spectral properties it is also understood how to formulate a consistent
quantum mechanical description for such non-Hermitian Hamiltonian systems 
\cite{Urubu,Mostafazadeh:2001nr,Bender:2002vv} by demanding the $\eta ^{2}$%
-operator to be Hermitian and positive-definite, such that it can be
interpreted as a metric to define the $\eta $-inner product. A special case
of this is the $\mathcal{CPT}$-inner product \cite{Bender:2002vv}, which
results by taking $\eta ^{2}=\mathcal{CP}$ with $\mathcal{C}=\sum \left\vert
\phi _{n}\right\rangle \left\langle \phi _{n}\right\vert $. For some recent
reviews on pseudo Hermitian Hamiltonians see \cite%
{specialCzech,special,CArev,Bendrev,special2}.

Since the metric-operator $\eta ^{2}$ is of central importance many attempts
have been made to construct it when given only a non-Hermitian Hamiltonian.
However, so far one has only succeeded to compute exact expressions for the
metric and isospectral partners in very few cases. Of course when the entire
spectrum is known this task is straightforward, even though one might not
always succeed to carry out the sum over all eigenfunctions. However, this
is a very special setting as even in the most simple cases one usually does
not have all the eigenfunctions at ones disposal and one has to resort to
more pragmatic techniques, such as for instance perturbation theory \cite%
{Bender:2004sa,Mostafazadeh:2004qh,CA,Can}. Rather than solving equations
for operators, the entire problem simplifies considerably if one converts it
into differential equations using Moyal products \cite{Moyal1,Moyal2,ACIso}
or other types of techniques \cite{MOT}. Here we wish to pursue the former
method for the most generic $\mathcal{PT}$-symmetric non-Hermitian
Hamiltonian of cubic order in the creation and annihilation operators.

We refer models for which the metric can be constructed exactly as \textit{%
solvable pseudo-Hermitian} (SPH) systems.

Our manuscript is organised as follows: In section 2 we introduce the model
we wish to investigate in this manuscript, formulating it in terms of
creation and annihilation operators and equivalently in terms of space and
momentum operators. We comment on the reduction of the model to models
previously studied. In section 3 we discuss in detail the method we are
going to employ to solve the equations (\ref{1}), namely to exploit the
isomorphism between products of operator valued functions and Moyal products
of scalar functions. In section 4 we construct systematically various exact
solutions for the metric operator and the Hermitian counterpart to $H$. As
special cases of these general considerations we focus in section 5 and 6 on
the single site lattice Reggeon model and the massive $\pm x^{6}$-potential.
In section 7 we provide a simple proof of the reality for the $ix^{2n+1}$%
-potentials and some of its generalizations. We state our conclusions in
section 8.

\section{A master Hamiltonian of cubic order}

The subject of our investigation is the most general $\mathcal{PT}$%
-symmetric Hamiltonian, which is maximally cubic in creation and
annihilation operators $a^{\dagger },a$, respectively, 
\begin{equation}
H_{\text{c}}={{\lambda }_{1}}a^{\dagger }a+{{\lambda }_{2}}a^{\dagger
}a^{\dagger }+{{\lambda }_{3}}aa+{{\lambda }_{4}+}i({{\lambda }_{5}}%
a^{\dagger }+{{\lambda }_{6}}a+{{\lambda }_{7}}a^{\dagger }a^{\dagger
}a^{\dagger }+{{\lambda }_{8}}a^{\dagger }a^{\dagger }a+{{\lambda }_{9}}%
a^{\dagger }aa+{{\lambda }_{10}}aaa).  \label{H211}
\end{equation}
The Hamiltonian $H_{\text{c}}$ is a ten-parameter family with $\lambda
_{i}\in \mathbb{R}$. It is clear that this Hamiltonian is $\mathcal{PT}$%
-symmetric by employing the usual identification $a=(\omega \hat{x}+i\hat{p}%
)/\sqrt{2\omega }$ and $a^{\dagger }=(\omega \hat{x}-i\hat{p})/\sqrt{2\omega 
}$ with the operators in $x$-space $\hat{x}$ and $\hat{p}=-i\partial _{x}$.
The effect of a simultaneous parity transformation $\mathcal{P}:\hat{x}%
\rightarrow -\hat{x}$ and time reversal $\mathcal{T}:t\rightarrow -t$, $%
i\rightarrow -i$ on the creation and annihilation operators is $\mathcal{PT}%
:a\rightarrow -a$, $a^{\dagger }\rightarrow -a^{\dagger }$. Without loss of
generality we may set the parameter $\omega $ to one in the following as it
is simply an overall energy scale.

In terms of the operators $\hat{x}$ and $\hat{p}$ the separation into a
Hermitian and non-Hermitian part is somewhat more transparent and we may
introduce in addition a coupling constant $g\in \mathbb{R}$ in order to be
able to treat the imaginary part as perturbation of a Hermitian operator. In
terms of these operators the most general expression is, as to be expected,
yet again a ten-parameter family 
\begin{equation}
H_{\text{c}}={{\alpha }_{1}}\hat{p}^{3}+{{\alpha }_{2}}\hat{p}^{2}+\alpha {%
_{3}}\frac{{\{}\hat{p},\hat{x}^{2}\}}{2}+\alpha {_{4}}\hat{p}+\alpha {_{5}}%
\hat{x}^{2}+\alpha {_{6}}+ig\left[ \alpha {_{7}}\frac{{\{}\hat{p}^{2},\hat{x}%
\}}{2}+\alpha {_{8}}\frac{{\{}\hat{p},\hat{x}\}}{2}+\alpha {_{9}}\hat{x}%
^{3}+\alpha {_{10}}\hat{x}\right] {.}  \label{Hc}
\end{equation}

\begin{center}
\begin{tabular}{|l||c|c|c|c|c|c|c|c|c|c|}
\hline
model\TEXTsymbol{\backslash}constants & $\alpha _{1}$ & $\alpha _{2}$ & $%
\alpha _{3}$ & $\alpha _{4}$ & $\alpha _{5}$ & $\alpha _{6}$ & $\alpha _{7}$
& $\alpha _{8}$ & $\alpha _{9}$ & $\alpha _{10}$ \\ \hline\hline
massive ix-potential & \multicolumn{1}{|c|}{$0$} & $1$ & $0$ & $0$ & $m^{2}$
& $0$ & $0$ & $0$ & $0$ & $1$ \\ \hline
massive ix$^{3}$-potential & \multicolumn{1}{|c|}{$0$} & $1$ & $0$ & $0$ & $%
m^{2}$ & $0$ & $0$ & $0$ & $1$ & $0$ \\ \hline
Swanson model & \multicolumn{1}{|c|}{$0$} & $\frac{\Delta }{2}$ & $0$ & $0$
& $\frac{\Delta }{2}$ & -$\frac{\Delta }{2}$ & $0$ & $1$ & $0$ & $0$ \\ 
\hline
lattice Reggeon & \multicolumn{1}{|c|}{$0$} & $\frac{\Delta }{2}$ & $0$ & $0$
& $\frac{\Delta }{2}$ & -$\frac{\Delta }{2}$ & $1$ & $0$ & $1$ & -$2$ \\ 
\hline
$\Delta a^{\dagger }a$+$iga^{\dagger }\left( a^{\dagger }\text{-}a\right) a$
& $g$ & $\frac{\Delta }{2}$ & $g$ & $-2g$ & $\frac{\Delta }{2}$ & -$\frac{%
\Delta }{2}$ & $0$ & $0$ & $0$ & $0$ \\ \hline
$H_{(\ref{EX1})}$ & $g$ & $\frac{\Delta }{2}$ & $g$ & $-2g$ & $\frac{\Delta 
}{2}$ & -$\frac{\Delta }{2}$ & $1$ & $0$ & $1$ & -$2$ \\ \hline
$H_{(\ref{SS3})}$ & \multicolumn{1}{|c|}{$0$} & $\frac{\Delta }{2}$ & $0$ & $%
0$ & $\frac{\Delta }{2}$ & -$\frac{\Delta }{2}$ & $1$ & $0$ & $0$ & -$2$ \\ 
\hline
$\frac{\hat{p}_{z}^{2}}{2}$-$\frac{g}{32}\hat{z}^{4}$ & \multicolumn{1}{|c|}{%
$0$} & $\frac{1}{2}$ & $0$ & $\frac{1}{4}-\frac{1}{2g}$ & $\frac{g}{2}$ & -$%
\frac{g}{2}$ & $\frac{1}{2g}$ & $0$ & $0$ & -$1$ \\ \hline
$\frac{\hat{p}_{z}^{2}}{2}$+$\lambda _{1}\hat{z}^{6}$+$\lambda _{2}\hat{z}%
^{2}$ & \multicolumn{1}{|c|}{$0$} & $\frac{1}{2}$ & $0$ & $\frac{1}{4}-\frac{%
1}{2g}$ & $192\lambda _{1}$ & $\kappa _{1}$ & $\frac{1}{2g}$ & $0$ & $\frac{%
64\lambda _{1}}{g}$ & $\frac{\kappa _{2}}{g}$ \\ \hline
\end{tabular}
\end{center}

\noindent {\small Table 1: Special reductions of the Hamiltonian $H_{\text{c}%
}.$ The map }$z(x)${\small \ is defined in equation (\ref{xz}) and $g,\alpha
,\Delta ,m\in \mathbb{R}$ are coupling constants of the models. We
abbreviated $\kappa _{1}=-4(16\lambda _{1}+\lambda _{2})$ and $\kappa
_{2}=-4(48\lambda _{1}+\lambda _{2})$.}

We have symmetrized in $H_{\text{c}}$ terms which contain $\hat{p}$ and $%
\hat{x}$ by introducing anticommutators, i.e. $\{A,B\}=AB+BA$. This allows
us to separate off conveniently the real and imaginary parts of $H_{\text{c}%
} $ by defining 
\begin{equation}
H_{\text{c}}(\hat{x},\hat{p})=h_{0}(\hat{x},\hat{p})+igh_{1}(\hat{x},\hat{p}%
),
\end{equation}%
with $h_{0}^{\dagger }=h_{0}$ and $h_{1}^{\dagger }=h_{1}$. In addition, the
symmetrized version (\ref{Hc}) will lead to very simple expressions when we
convert products of operator valued functions into expressions involving
scalar functions multiplied via Moyal products. For our definition of the
Moyal product it implies that the parameters $\alpha $ do not need to be
re-defined. Depending on the context, one (\ref{H211}) or the other (\ref{Hc}%
) formulation is more advantageous. Whereas the usage of creation and
annihilation operators is more proned for an algebraic generalization, see
e.g. \cite{Quesne}, the formulation in terms of operators $\hat{x}$ and $%
\hat{p}$ is more suitable for a treatment with Moyal brackets. The relation
between the two versions is easily computed from the aforementioned
identifications between the $a$,$a^{\dagger }$ and $\hat{x}$,$\hat{p}~$via
the relations $\alpha {{=M}\lambda }$ and $\lambda =M^{-1}\alpha $ with $M$
being a $10\times 10$-matrix. Below we will impose some constraints on the
coefficients $\alpha $ and it is therefore useful to have an explicit
expression for $M$ at our disposal in order to see how these constraints
affect the expression for the Hamiltonian in (\ref{H211}). We compute the
matrix 
\begin{equation}
M=\left( 
\begin{array}{cccccccccc}
0 & 0 & 0 & 0 & 0 & 0 & -\frac{1}{2\sqrt{2}} & \frac{1}{2\sqrt{2}} & -\frac{1%
}{2\sqrt{2}} & \frac{1}{2\sqrt{2}} \\ 
\frac{1}{2} & -\frac{1}{2} & -\frac{1}{2} & 0 & 0 & 0 & 0 & 0 & 0 & 0 \\ 
0 & 0 & 0 & 0 & 0 & 0 & \frac{3}{2\sqrt{2}} & \frac{1}{2\sqrt{2}} & -\frac{1%
}{2\sqrt{2}} & -\frac{3}{2\sqrt{2}} \\ 
0 & 0 & 0 & 0 & \frac{1}{\sqrt{2}} & -\frac{1}{\sqrt{2}} & 0 & -\frac{1}{%
\sqrt{2}} & \frac{1}{\sqrt{2}} & 0 \\ 
\frac{1}{2} & \frac{1}{2} & \frac{1}{2} & 0 & 0 & 0 & 0 & 0 & 0 & 0 \\ 
-\frac{1}{2} & 0 & 0 & 1 & 0 & 0 & 0 & 0 & 0 & 0 \\ 
0 & 0 & 0 & 0 & 0 & 0 & -\frac{3}{2\sqrt{2}g} & \frac{1}{2\sqrt{2}g} & \frac{%
1}{2\sqrt{2}g} & -\frac{3}{2\sqrt{2}g} \\ 
0 & -\frac{1}{g} & \frac{1}{g} & 0 & 0 & 0 & 0 & 0 & 0 & 0 \\ 
0 & 0 & 0 & 0 & 0 & 0 & \frac{1}{2\sqrt{2}g} & \frac{1}{2\sqrt{2}g} & \frac{1%
}{2\sqrt{2}g} & \frac{1}{2\sqrt{2}g} \\ 
0 & 0 & 0 & 0 & \frac{1}{g\sqrt{2}} & \frac{1}{g\sqrt{2}} & 0 & -\frac{1}{g%
\sqrt{2}} & -\frac{1}{g\sqrt{2}} & 0%
\end{array}%
\right)
\end{equation}%
and the inverse 
\begin{equation}
M^{-1}=\left( 
\begin{array}{cccccccccc}
0 & 1 & 0 & 0 & 1 & 0 & 0 & 0 & 0 & 0 \\ 
0 & -\frac{1}{2} & 0 & 0 & \frac{1}{2} & 0 & 0 & -\frac{g}{2} & 0 & 0 \\ 
0 & -\frac{1}{2} & 0 & 0 & \frac{1}{2} & 0 & 0 & \frac{g}{2} & 0 & 0 \\ 
0 & \frac{1}{2} & 0 & 0 & \frac{1}{2} & 1 & 0 & 0 & 0 & 0 \\ 
\frac{3}{2\sqrt{2}} & 0 & \frac{1}{2\sqrt{2}} & \frac{1}{\sqrt{2}} & 0 & 0 & 
-\frac{g}{2\sqrt{2}} & 0 & \frac{3\,g}{2\sqrt{2}} & \frac{g}{\sqrt{2}} \\ 
-\frac{3}{2\sqrt{2}} & 0 & -\frac{1}{2\sqrt{2}} & -\frac{1}{\sqrt{2}} & 0 & 0
& \frac{g}{2\sqrt{2}} & 0 & \frac{3\,g}{2\sqrt{2}} & \frac{g}{\sqrt{2}} \\ 
-\frac{1}{2\sqrt{2}} & 0 & \frac{1}{2\sqrt{2}} & 0 & 0 & 0 & -\frac{g}{2%
\sqrt{2}} & 0 & \frac{g}{2\sqrt{2}} & 0 \\ 
\frac{3}{2\sqrt{2}} & 0 & \frac{1}{2\sqrt{2}} & 0 & 0 & 0 & \frac{g}{2\sqrt{2%
}} & 0 & \frac{3\,g}{2\sqrt{2}} & 0 \\ 
-\frac{3}{2\sqrt{2}} & 0 & -\frac{1}{2\sqrt{2}} & 0 & 0 & 0 & \frac{g}{2%
\sqrt{2}} & 0 & \frac{3\,g}{2\sqrt{2}} & 0 \\ 
\frac{1}{2\sqrt{2}} & 0 & -\frac{1}{2\sqrt{2}} & 0 & 0 & 0 & -\frac{g}{2%
\sqrt{2}} & 0 & \frac{g}{2\sqrt{2}} & 0%
\end{array}%
\right) ,
\end{equation}%
which also exists for $g\neq 0$ since $\det M=-g^{-4}$.

The Hamiltonian $H_{\text{c}}$ encompasses many models and for specific
choices of some of the $\alpha _{i}$ it reduces to various well studied
examples, such as the simple massive ix-potential \cite{Bender:2005sc} or
its massless version, the so-called Swanson Hamiltonian \cite%
{Swanson,HJ,JM,ACIso,MGH}, the complex cubic potential together with his
massive version \cite{Bender:1998ke} and also the transformed version of the 
$-\hat{x}^{4}$-potential \cite{JM}. As we will show below, in addition it
includes several interesting new models, such as the single site lattice
version of Reggeon field theory \cite{Regge2}, which is a thirty year old
model but has not been considered in the current context and the transformed
version of the $\pm x^{6}$- potential, which serves as a toy model to
identify theories with vanishing cosmological constant \cite{thooft}. The
latter models have not been solved so far with regard to their metric
operators and isospectral partners. Besides these models, $H_{\text{c}}$
also includes many new models not considered so far, some of which are even
SPH.

To enable easy reference we summarize the various choices in table 1.

Most SPH-models which have been constructed so far are rather trivial, such
as the massive $ix$-potential or the so-called Swanson Hamiltonian. The
latter model can be obtained simply from the standard harmonic oscillator by
means of a Bogolyubov transformation and a subsequent similarity
transformation, which is bilinear in $a$ and $a^{\dagger }$. Beyond these
maximally quadratic models, the complex cubic potential was the first model
which has been studied in more detail. Unfortunately so far it can only be
treated perturbatively. The transformed version of the from below unbounded
-z$^{4}$-potential is the first SPH-model containing at least one cubic
term. Here we enlarge this class of models. As a special case we shall also
investigate the single site lattice version of Reggeon field theory \cite%
{Regge2} in more detail. Before treating these specific models let us
investigate first the Hamiltonian $H_{\text{c}}$ in a very generic manner.

Our objective is to solve equation (\ref{1}) and find an exact expression
for the positive-definite metric operator $\eta ^{2}$, subsequently to solve
for the similarity transformation and construct Hermitian isospectral
partner Hamiltonians.

\section{Pseudo-Hermitian Hamiltonians from Moyal products}

\subsection{Generalities}

Taking solely a non-Hermitian Hamiltonian as a starting point, there is of
course not a one-to-one correspondence to one specific Hermitian Hamiltonian
counterpart. The conjugation relation in (\ref{1}) admits obviously a whole
family of solutions. In order to construct these solutions we will not use
commutation relations involving operators, but instead we will exploit the
isomorphism between operator valued function in $\hat{x}$ and $\hat{p}$ and
scalar functions multiplied by Moyal products in monomial of scalars $x$ and 
$p$. We associate to two arbitrary operator valued functions $F(\hat{x},\hat{%
p})$ and $G(\hat{x},\hat{p})$ two scalar functions $F(x,p)$, $G(x,p)\in 
\mathcal{S}$ such that 
\begin{equation}
F(\hat{x},\hat{p})G(\hat{x},\hat{p})\cong F(x,p)\mathcal{\star }G(x,p),
\label{iso}
\end{equation}
where $\mathcal{S}$ is the space of complex valued integrable functions.
Here we use the following standard definition of the Moyal product $\star :%
\mathcal{S\times S\rightarrow S}$, see e.g. \cite%
{Fairlie:1998rf,Carroll,ACIso}, 
\begin{align}
F(x,p)\star G(x,p)& =F(x,p)e^{\frac{i}{2}(\overleftarrow{{\partial }}_{\!\!x}%
\overrightarrow{{\partial }}_{\!\!p}-\overleftarrow{{\partial }}_{\!\!p}%
\overrightarrow{{\partial }}_{\!\!x})}G(x,p)  \label{star} \\
& =\sum\limits_{s=0}^{\infty }\frac{(-i/2)^{s}}{s!}\sum%
\limits_{t=0}^{s}(-1)^{t}\left( 
\begin{array}{r}
s \\ 
t%
\end{array}
\right) \partial _{x}^{t}\partial _{p}^{s-t}F(x,p)\partial
_{x}^{s-t}\partial _{p}^{t}G(x,p).  \notag
\end{align}
The Moyal product is a distributive and associative map obeying the same
Hermiticity properties as the operator valued functions on the right hand
side of (\ref{iso}), that is $(F\star G)^{\ast }=G^{\ast }\star F^{\ast }$.
Following standard arguments we provide now explicit representations for the 
$F(\hat{x},\hat{p})$ and $F(x,p)$. We may formally Fourier expand an
arbitrary operator valued functions $F(\hat{x},\hat{p})$ and scalar
functions $F(x,p)$ as 
\begin{equation}
F(\hat{x},\hat{p})=\int\nolimits_{-\infty }^{\infty }dsdtf(s,t)e^{i(s\hat{x}%
+t\hat{p})}\quad \text{and\quad }F(x,p)=\int\nolimits_{-\infty }^{\infty
}dsdtf(s,t)e^{i(sx+tp)},  \label{rep}
\end{equation}
respectively. In terms of this representation the multiplication of two
operator valued functions yields 
\begin{equation}
F(\hat{x},\hat{p})G(\hat{x},\hat{p})=\int\nolimits_{-\infty }^{\infty
}dsdtds^{\prime }dt^{\prime }f(s,t)f(s^{\prime },t^{\prime })e^{\frac{i}{2}%
(ts^{\prime }-t^{\prime }s)}e^{i(s+s^{\prime })\hat{x}+i(t+t^{\prime })\hat{p%
}},  \label{FG}
\end{equation}
which follows using the identities $e^{i(s\hat{x}+t\hat{p})}=e^{is\hat{x}%
/2}e^{it\hat{p}}e^{is\hat{x}/2}$ and $e^{is\hat{x}/2}e^{it\hat{p}}=e^{it\hat{%
p}}e^{is\hat{x}/2}e^{ist}$. Is now straightforward to verify that the
definition of the Moyal product (\ref{star}) guarantees that the isomorphism
(\ref{iso}) holds, since $F(x,p)\mathcal{\star }G(x,p)$ yields formally the
same expression as (\ref{FG}) with $\hat{x},\hat{p}$ replaced by $x,p$.

The Hermiticity property is important for our purposes. We find that 
\begin{equation}
F^{\dagger }(\hat{x},\hat{p})=F(\hat{x},\hat{p})\cong F^{\ast }(x,p)=F(x,p).
\label{Herm}
\end{equation}%
This is easily seen by computing $F^{\dagger }(\hat{x},\hat{p})$ using the
representation (\ref{rep}). Then this function is Hermitian if and only if
the kernel satisfies $f^{\ast }(s,t)=f(-s,-t)$, which in turn implies that $%
F(x,p)$ is real. Positive definiteness of an operator valued function $F(%
\hat{x},\hat{p})$ is guaranteed if the logarithm of the operator is
Hermitian, that is we need to ensure that $\log F(x,p)$ is real.
Furthermore, it is easy to see that $F(\hat{x},\hat{p})$ is $\mathcal{PT}$%
-symmetric if and only if $f^{\ast }(s,t)=f(s,-t)$.

As an instructive example we consider  $F(x,p)=x^{m}p^{n}$ for which we
compute the corresponding kernel as $f(s,t)=i^{n+m}\delta ^{(m)}(s)\delta
^{(n)}(t)$. From this it is easy to see that $(ix)^{m}p^{n}$ is $\mathcal{PT}
$-symmetric, since$\ i^{m}f(s,t)$ satisfies $\left[ i^{m}f(s,t)\right]
^{\ast }=i^{m}f(s,-t)$.

In the present context of studying non-Hermitian Hamiltonians this technique
of exploiting the isomorphism between Moyal products and operator products
has been exploited by Scholtz and Geyer \cite{Moyal1,Moyal2}, who reproduced
some previously known results and also in \cite{ACIso}, where new solutions
were constructed. In \cite{Moyal1,Moyal2} a more asymmetrical definition
than (\ref{star}) of the Moyal product was employed, i.e. $F(x,p)\ast
G(x,p)=F(x,p)e^{i\overleftarrow{{\partial }}_{\!\!x}\overrightarrow{{%
\partial }}_{\!\!p}}G(x,p)$. In comparison with (\ref{star}) this definition
leads to some rather unappealing properties: i) the loss of the useful and
natural Hermiticity relation, i.e. $(F\ast G)^{\ast }\neq G^{\ast }\ast
F^{\ast }$, ii) the right hand side of the isomorphism in (\ref{Herm}) is
replaced by the less transparent expression $F^{\ast }(x,p)=e^{-i{\partial }%
_{x}{\partial }_{p}}F(x,p)$ and iii) in \cite{ACIso} it was shown that the
definition $\ast $ leads to more complicated differential equations than the
definition $\star $. The representation for the operator valued functions $F(%
\hat{x},\hat{p})$, which satisfies the properties resulting from the
definition $\ast $ differs from (\ref{rep}) by replacing $e^{i(s\hat{x}+t%
\hat{p})}\rightarrow e^{is\hat{x}}e^{it\hat{p}}$ in the Fourier expansion.

\subsection{Construction of the metric operator and isospectral partners}

We briefly recapitulate the main steps of the procedure \cite%
{Moyal1,Moyal2,ACIso} of how to find for a given non-Hermitian Hamiltonian $%
H $ a metric operator $\eta ^{2}(\hat{x},\hat{p})$, a similarity
transformation $\eta (\hat{x},\hat{p})$ and an Hermitian counterpart $h(\hat{%
x},\hat{p})$ using Moyal products. First of all we need to solve the right
hand side of the isomorphism 
\begin{equation}
H^{\dagger }(\hat{x},\hat{p})\eta ^{2}(\hat{x},\hat{p})=\eta ^{2}(\hat{x},%
\hat{p})H(\hat{x},\hat{p})\cong H^{\dagger }(x,p)\mathcal{\star }\eta
^{2}(x,p)=\eta ^{2}(x,p)\mathcal{\star }H(x,p)  \label{Het}
\end{equation}
for the ``scalar metric function''\ $\eta ^{2}(x,p)$. Taking as a starting
point the non-Hermitian Hamiltonian $H(\hat{x},\hat{p})$, we have to
transform this expression into a scalar function $H(x,p)$ by replacing all
occurring operator products with Moyal products. We can use this expression
to evaluate the right hand side of the isomorphism of (\ref{Het}), which is
a differential equation for $\eta ^{2}(x,p)$ whose order is governed by the
highest powers of $x$ and $p$ in $H(x,p)$. Subsequently we may replace the
function $\eta ^{2}(x,p)$ by the metric operator $\eta ^{2}(\hat{x},\hat{p})$
using the isomorphism (\ref{iso}) now in reverse from the right to the left.
Thereafter we solve the differential equation $\eta ^{2}(x,p)=\eta
(x,p)\star \eta (x,p)$ for $\eta (x,p)$. Inverting this expression we obtain 
$\eta ^{-1}(x,p)$, such that we are equipped to compute directly the scalar
function associated to the Hermitian counterpart by evaluating 
\begin{equation}
h(x,p)=\eta (x,p)\star H(x,p)\star \eta ^{-1}(x,p).  \label{hH}
\end{equation}
Finally we have to convert the function $\eta (x,p)$ into the operator
valued function $\eta (\hat{x},\hat{p})$ and the ``Hermitian scalar
function''\ $h(x,p)$ into the Hamiltonian counterpart $h(\hat{x},\hat{p})$,
once more by solving (\ref{iso}) from the right to the left.

So far we did not comment on whether the metric is a meaningful Hermitian
and positive operator. According to the isomorphism (\ref{Herm}) we simply
have to verify that $\eta ^{2}(x,p)$, $\eta (x,p)$ and $h(x,p)$ are real
functions in order to establish that the corresponding operator valued
functions $\eta ^{2}(\hat{x},\hat{p})$, $\eta (\hat{x},\hat{p})$ and $h(\hat{%
x},\hat{p})$ are Hermitian. We may establish positive definiteness of these
operators by verifying that their logarithms are real.

\subsection{Ambiguities in the solution}

Obviously when having a non-Hermitian Hamiltonian as the sole starting point
there is not a unique Hermitian counterpart in the same similarity class
associated to the adjoint action of one unique operator $\eta $.
Consequently also the metric operator $\eta ^{2}$ is not unique. The latter
was pointed out for instance in \cite{Moyal2} and exemplified in detail for
the concrete example of the so-called Swanson Hamiltonian in \cite{MGH}. In
fact, it is trivial to see that any two non-equivalent metric operators, say 
$\eta ^{2}$ and $\hat{\eta}^{2}$, can be used to construct a non-unitary
symmetry operator $S:=\eta ^{-2}\hat{\eta}^{2}$ $\neq S^{\dagger }=\hat{\eta}%
^{2}\eta ^{-2}$ for the non-Hermitian Hamiltonian $H$ 
\begin{equation}
H^{\dagger }=\eta ^{2}H\eta ^{-2}=\hat{\eta}^{2}H\hat{\eta}^{-2}\qquad
\Leftrightarrow \qquad \lbrack S,H]=[S^{\dagger },H^{\dagger }]=0.
\label{symm}
\end{equation}
We may solve (\ref{symm}) and express one metric in terms of the other as 
\begin{equation}
\hat{\eta}^{2}=\left( H^{\dagger }\right) ^{n}\eta ^{2}H^{n}\text{\qquad for 
}n\in \mathbb{N}.
\end{equation}
Thus we encounter here an infinite amount of new solutions. Likewise this
ambiguity can be related to the non-equivalent Hermitian counterparts 
\begin{equation}
h=\eta H\eta ^{-1},\hat{h}=\hat{\eta}H\hat{\eta}^{-1}\qquad \Leftrightarrow
\qquad \lbrack s,h]=[\hat{s},\hat{h}]=0,  \label{S2}
\end{equation}
with symmetry operators $s=\eta \hat{\eta}^{-2}\eta $ and $\hat{s}=\hat{\eta}%
^{-1}\eta ^{2}\hat{\eta}^{-1}$. When $\eta ^{\dagger }=\eta $ and $\hat{\eta}%
^{\dagger }=\hat{\eta}$ we obviously also have $s^{\dagger }=s$ and $\hat{s}%
^{\dagger }=\hat{s}$. The expression for the symmetry operator $s$ for $h$
was also identified in \cite{Most}.

There are various ways to select a unique solution. One possibility \cite%
{Urubu} is to specify one more observable in the non-Hermitian system.
However, this argument is very impractical as one does not know a priori
which variables constitute observables.

\section{SPH-models of cubic order}

Let us study $H_{\text{c}}(\hat{x},\hat{p})$ by converting it first into a
scalar function $H_{\text{c}}(x,p)$. Most terms are non problematic and we
can simply substitute $\hat{x}\rightarrow x,\hat{p}\rightarrow p$, but
according to our definition of the Moyal bracket (\ref{star}) we have to
replace $\hat{p}^{2}\hat{x}\rightarrow p^{2}\star x=p^{2}x-ip$, $\hat{p}\hat{%
x}^{2}\rightarrow p\star x^{2}=px^{2}-ix$, $\hat{p}\hat{x}\rightarrow p\star
x=px-i/2$ etc. Replacing all operator products in this way we convert the
Hamiltonian $H_{\text{c}}(\hat{x},\hat{p})$ in (\ref{Hc}) into the scalar
function 
\begin{equation}
H_{\text{c}}(x,p)=\alpha {_{1}}p^{3}+\alpha {_{2}}p^{2}+\alpha {_{3}}%
px^{2}+\alpha {_{4}}p+\alpha {_{5}}x^{2}+\alpha {_{6}}+ig(\alpha {_{7}}%
p^{2}x+\alpha {_{8}}px+\alpha {_{9}}x^{3}+\alpha {_{10}}x\,{).}  \label{hcs}
\end{equation}%
Substituting (\ref{hcs}) into the right hand side of the isomorphism into (%
\ref{Het}) yields the third order differential equation 
\begin{eqnarray}
&&\left( {{\alpha }_{3}}px{\partial }_{p}+{{\alpha }_{5}}x{\partial }_{p}+%
\frac{{{\alpha }_{3}}}{8}{\partial }_{x}{\partial }_{p}^{2}+\frac{{{\alpha }%
_{1}}}{8}{\partial }_{x}^{3}-{{\alpha }_{2}}p{\partial }_{x}-\frac{3}{2}{{%
\alpha }_{1}}p^{2}{\partial }_{x}-\frac{{{\alpha }_{3}}}{2}x^{2}{\partial }%
_{x}-\frac{{{\alpha }_{4}}}{2}{\partial }_{x}\right) \eta ^{2}~~~~~~
\label{eq} \\
&=&g\left( {{\alpha }_{9}}x^{3}+{{\alpha }_{10}}x+{{\alpha }_{8}}px+{{\alpha 
}_{7}}p^{2}x+\frac{{{\alpha }_{7}}}{2}p{\partial }_{x}{\partial }_{p}+\frac{{%
{\alpha }_{8}}}{4}{\partial }_{x}{\partial }_{p}-\frac{{{\alpha }_{7}}}{4}x{%
\partial }_{x}^{2}-\frac{3}{4}{{\alpha }_{9}}x{\partial }_{p}^{2}\right)
\eta ^{2}  \notag
\end{eqnarray}%
for the \textquotedblleft metric scalar function\textquotedblright\ $\eta
^{2}(x,p)$. There are various simplifications one can make at this stage.
First of all we could assume that either $\hat{x}$ or $\hat{p}$ is an
observable in the non-Hermitian system, such that $\eta ^{2}(x,p)$ does not
depend on $p$ or $x$, respectively. As pointed out before it is not clear at
this stage if any of these choices is consistent. However, any particular
choice $p$ or $x$ will be vindicated if (\ref{eq}) can be solved
subsequently for $\eta ^{2}(p)$ or $\eta ^{2}(x)$, respectively. Here we
will assume that $\eta ^{2}(x,p)$ admits a perturbative expansion. Making a
very generic exponential $\mathcal{PT}$-symmetric ansatz, which is real and
cubic in its argument for $\eta ^{2}(x,p)=\exp
g(q_{1}p^{3}+q_{2}px^{2}+q_{3}p^{2}+q_{4}x^{2}+q_{5}p)$, we construct
systematically all exact solutions of this form. Substituting the ansatz
into the differential equation (\ref{eq}) and reading off the coefficients
in front of each monomial in $x$ and $p$ yields at each order in $g$ ten
equations. by solving these equations we find five qualitatively different
types of exact solutions characterized by vanishing coefficients $\alpha
_{i} $ and some additional constraints. We will now present these solutions.

\subsection{ Non-vanishing $\hat{p}\hat{x}^{2}$-term}

\subsubsection{Constraints 1}

We consider the full Hamiltonian $H_{\text{c}}(x,p)$ in (\ref{hcs}) and
impose as the only constraint that the $px^{2}$-term does not vanish, i.e. ${%
{\alpha }_{3}\neq 0}$. For this situation we can solve the differential
equation (\ref{eq}) exactly to all orders in perturbation theory for 
\begin{equation}
H_{\text{c}}(x,p)=h_{0}(x,p)+ig(\frac{{{\alpha }_{1}{\alpha }_{9}}}{{{\alpha 
}_{3}}}p^{2}x+\frac{{{\alpha }_{2}{\alpha }_{9}-{\alpha }_{5}{\alpha }_{7}}}{%
{{\alpha }_{3}}}px+\alpha {_{9}}x^{3}+\frac{{{\alpha }_{4}{\alpha }_{9}-{%
\alpha }_{5}{\alpha }_{8}}}{{{\alpha }_{3}}}x\,{),}  \label{HH}
\end{equation}%
where we imposed the additional constraints 
\begin{equation}
{{\alpha }_{1}{\alpha }_{9}={\alpha }_{3}{\alpha }_{7},\qquad {\alpha }_{2}{%
\alpha }_{9}={\alpha }_{5}{\alpha }_{7}+{\alpha }_{3}{\alpha }_{8}\qquad }%
\text{and\qquad }{{\alpha }_{4}{\alpha }_{9}={\alpha }_{5}{\alpha }_{8}+{%
\alpha }_{3}{\alpha }_{10}.}  \label{con1}
\end{equation}%
In (\ref{HH}) we have replaced the constants ${{\alpha }_{7},{\alpha }_{8}}$
and ${{\alpha }_{10}}$ using (\ref{con1}). The solution of the differential
equation is the metric scalar function 
\begin{equation}
\eta ^{2}(x,p)=e^{-g\left( \frac{{{\alpha }_{7}}}{{{\alpha }_{3}}}\,p^{2}+%
\frac{\,{{\alpha }_{8}}}{{{\alpha }_{3}}}p+\frac{{{\alpha }_{9}}}{{{\alpha }%
_{3}}}x^{2}\right) }.  \label{m1}
\end{equation}%
Since $\eta ^{2}(x,p)$ is real it follows from (\ref{Herm}) that the
corresponding metric operator is Hermitian. Next we solve $\eta (x,p)\star
\eta (x,p)=\eta ^{2}(x,p)$ for $\eta (x,p)$. Up to order $g^{2}$ we find 
\begin{eqnarray}
\eta (x,p) &=&1-g\frac{{{\alpha }_{7}}p^{2}+{{\alpha }_{8}}p+x^{2}{{\alpha }%
_{9}}}{2{{\alpha }_{3}}}+g^{2}\left( \frac{{{\alpha }_{9}}\left( {{\alpha }%
_{7}}+2{{\alpha }_{7}}p^{2}x^{2}+2{{\alpha }_{8}}px^{2}\right) +{{{\alpha }%
_{9}^{2}}}x^{4}}{8{{{\alpha }_{3}^{2}}}}\right. ~~~~  \notag \\
&&+\left. \frac{{\left( p\alpha {_{7}}+{{\alpha }_{8}}\right) }^{2}p^{2}}{8{{%
{\alpha }_{3}^{2}}}}\right) +\mathcal{O}(g^{3}).
\end{eqnarray}%
The corresponding Hermitian counterpart corresponding to this solution is
computed by means of (\ref{hH}) to 
\begin{eqnarray}
h_{\text{c}}(x,p) &=&{{\alpha }_{3}}px^{2}+{{\alpha }_{5}}x^{2}+{{\alpha }%
_{6}}+\frac{{{\alpha }_{3}{\alpha }_{7}}}{{{\alpha }_{9}}}p^{3}+\frac{\left( 
{{\alpha }_{5}{\alpha }_{7}}+{{\alpha }_{3}{\alpha }_{8}}\right) }{{{\alpha }%
_{9}}}p^{2}+\frac{\left( {{\alpha }_{5}{\alpha }_{8}}+{{\alpha }_{3}{\alpha }%
_{10}}\right) }{{{\alpha }_{9}}}p~~~~  \notag \\
&&-g^{2}\frac{\left( 2\,{{\alpha }_{7}}p\,+{{\alpha }_{8}}\right) \,\left(
p\,\left( {{\alpha }_{7}}p\,+{{\alpha }_{8}}\right) +{{\alpha }_{9}}x^{2}\,+{%
{\alpha }_{10}}\right) }{4\,{{\alpha }_{3}}}+\mathcal{O}(g^{4}).
\end{eqnarray}%
Notice that since we demanded ${{\alpha }_{3}}$ to be non-vanishing these
solutions can not be reduced to any of the well studied models presented in
table 1, but represent new types of solutions. We may simplify the above
Hamiltonians by setting various $\alpha $s to zero.

Demanding for instance that $\hat{x}$ is an observable in the non-Hermitian
system we are forced by (\ref{m1}) to set ${{\alpha }_{7}={\alpha }_{8}=0}$
and by (\ref{con1}) also ${{\alpha }_{1}={\alpha }_{2}=0}$. The Hamiltonian
in (\ref{HH}) then simplifies to 
\begin{equation}
H_{\text{c}}(x,p)=\alpha {_{3}}px^{2}+\alpha {_{4}}p+\alpha {_{5}}%
x^{2}+\alpha {_{6}}+ig(\alpha {_{9}}x^{3}+\frac{{{\alpha }_{4}{\alpha }_{9}}%
}{{{\alpha }_{3}}}x\,{).}
\end{equation}
Since $\eta ^{2}(x,p)$ only depends on $x$ in this case, we can compute
exactly $\eta (x,p)=e^{-g\frac{{{\alpha }_{9}}}{2{{\alpha }_{3}}}x^{2}}$.
The Hermitian counterpart results to 
\begin{equation}
h_{\text{c}}(x,p)=h_{0}(x,p)=\alpha {_{3}}px^{2}+\alpha {_{4}}p+\alpha {_{5}}%
x^{2}+\alpha {_{6}.}
\end{equation}

If we require on the other hand that $\hat{p}$ is an observable, we have to
choose ${{\alpha }_{9}\rightarrow 0}$. However, in that case the constraints
(\ref{con1}) imply that the non-Hermitian part of the Hamiltonian (\ref{HH})
vanishes, i.e. we obtain the trivial case $H_{\text{c}}(x,p)=h_{0}(x,p)$.

\subsubsection{Constraints 2}

In the construction of the previous solution some coefficients had to
satisfy a quadratic equations in the parameters to guarantee the vanishing
of the perturbative expansion. The other solution for this equation leads to
the constraints ${{\alpha }_{1}={\alpha }_{7}=0}$, such that the
non-Hermitian Hamiltonian simplifies. If we now impose the additional
constraints 
\begin{equation}
{{\alpha }_{3}{\alpha }_{10}={\alpha }_{4}{\alpha }_{9}\qquad }\text{%
and\qquad }{{\alpha }_{3}{\alpha }_{8}=2{\alpha }_{2}{\alpha }_{9},}
\label{con2}
\end{equation}
we can solve the differential equation (\ref{eq}) exactly. For 
\begin{equation}
H_{\text{c}}(x,p)=\alpha {_{2}}p^{2}+\alpha {_{3}}px^{2}+\alpha {_{4}}%
p+\alpha {_{5}}x^{2}+\alpha {_{6}}+ig(\frac{{2{\alpha }_{2}{\alpha }_{9}}}{{{%
\alpha }_{3}}}px+\alpha {_{9}}x^{3}+\frac{{{\alpha }_{4}{\alpha }_{9}}}{{{%
\alpha }_{3}}}x\,{)}  \label{H2}
\end{equation}
we compute the exact scalar metric function to 
\begin{equation}
\eta ^{2}(x,p)=e^{-g\frac{{{\alpha }_{9}}}{{{\alpha }_{3}}}x^{2}}.
\end{equation}
Clearly $\eta ^{2}(\hat{x},\hat{p})$ is a Hermitian and positive definite
operator, which follows from the facts that $\eta ^{2}(x,p)$ and $\log \eta
^{2}(x,p)$ are real, respectively. Notice the fact that the Hamiltonian (\ref%
{H2}) does not follow as a specialization of (\ref{HH}), since the
constraints (\ref{con2}) do not result as a particular case of (\ref{con1}).
The Hermitian Hamiltonian counterpart corresponding \ to (\ref{H2}) is
computed with $\eta ^{2}(x,p)=e^{-g\frac{{{\alpha }_{9}}}{2{{\alpha }_{3}}}%
x^{2}}$by means of (\ref{hH}) to 
\begin{equation}
h_{\text{c}}(x,p)={{\alpha }_{2}}p^{2}\,+{{\alpha }_{3}}px^{2}\,+{{\alpha }%
_{4}}p\,+{{\alpha }_{5}}x^{2}\,+{{\alpha }_{6}}+g^{2}\frac{\,{{\alpha }_{2}}%
\,{{{\alpha }_{9}^{2}}}}{{{{\alpha }_{3}}}^{2}}\,x^{2}.
\end{equation}
Once again we may simplify the above Hamiltonians by setting various $\alpha 
$s to zero or other special values, except for the case ${{\alpha }%
_{9}\rightarrow 0}$ for which the constraints (\ref{con2}) reduce the
non-Hermitian part of the Hamiltonian (\ref{H2}) to zero.

Thus this case requires a separate consideration:

\subsection{Non-vanishing $\hat{p}\hat{x}^{2}$-term and vanishing $\hat{x}%
^{3}$-term}

Let us therefore embark on the treatment of the complementary case to the
previous subsection, namely ${{\alpha }_{3}\neq 0}$ and ${{\alpha }_{9}=0}$.
For these constraints we can solve the differential equation (\ref{eq})
exactly for the Hamiltonian 
\begin{equation}
H_{\text{c}}(x,p)=h_{0}(x,p)+ig(\alpha {_{7}}p^{2}x+\alpha {_{8}}px+\frac{{{%
\alpha }_{5}({\alpha }_{3}{\alpha }_{8}-{\alpha }_{5}{\alpha }_{7})}}{{{%
\alpha }_{3}^{2}}}x\,{),}  \label{H3}
\end{equation}%
when we impose one additional constraint 
\begin{equation}
{{\alpha }_{10}{\alpha }_{3}^{2}={\alpha }_{5}({\alpha }_{3}{\alpha }_{8}-{%
\alpha }_{5}{\alpha }_{7}).}  \label{xy}
\end{equation}%
The \textquotedblleft metric scalar function\textquotedblright\ results to 
\begin{equation}
\eta ^{2}(x,p)=\eta ^{2}(p)=e^{g\left( \frac{{{\alpha }_{7}}}{2{{\alpha }_{3}%
}}p^{2}+\frac{{{{\alpha }_{3}\alpha }_{8}}-\,{\alpha }_{5}{{\alpha }_{7}}}{{{%
{\alpha }_{3}}}^{2}}p\right) }.
\end{equation}%
Once again $\eta ^{2}(\hat{x},\hat{p})$ is a Hermitian and positive definite
operator, which follows again from the facts that $\eta ^{2}(x,p)$ and $\log
\eta ^{2}(x,p)$ are real. Since $\eta ^{2}(x,p)$ only depends on $p$, we can
simply take the square root to compute $\eta (p)$. Then the corresponding
Hermitian counterpart is computed by means of (\ref{hH}) to 
\begin{equation}
h_{\text{c}}={h}_{0}+g^{2}\left( \frac{{{{\alpha }_{7}}}^{2}}{4{{\alpha }_{3}%
}}p^{3}+\frac{2{{\alpha }_{3}{\alpha }_{7}{\alpha }_{8}}-{{\alpha }_{5}{%
\alpha }_{7}^{2}}}{4{{{\alpha }_{3}}}^{2}}p^{2}+\frac{{{{\alpha }_{3}}}^{2}{{%
{\alpha }_{8}}}^{2}-{{{\alpha }_{5}}}^{2}{{{\alpha }_{7}}}^{2}}{4{{{\alpha }%
_{3}}}^{3}}p+\frac{{{\alpha }_{5}\left( {{\alpha }_{5}{\alpha }_{7}}-{{%
\alpha }_{3}{\alpha }_{8}}\right) }^{2}}{4{{{\alpha }_{3}}}^{4}}\right)
\label{a22}
\end{equation}%
In fact we can implement the constraint (\ref{xy}) directly in the solution.
The function 
\begin{equation}
\eta ^{2}(p)={\left( p\,{{\alpha }_{3}}+{{\alpha }_{5}}\right) }^{\frac{%
g\,\left( {{{\alpha }_{5}}}^{2}\,{{\alpha }_{7}}-{{\alpha }_{3}}\,{{\alpha }%
_{5}}\,{{\alpha }_{8}}+{{{\alpha }_{3}}}^{2}\,{{\alpha }_{10}}\right) }{{{{%
\alpha }_{3}}}^{3}}}e^{g\left( \frac{{{\alpha }_{7}}}{2{{\alpha }_{3}}}p^{2}+%
\frac{{{{\alpha }_{3}\alpha }_{8}}-\,{\alpha }_{5}{{\alpha }_{7}}}{{{{\alpha 
}_{3}}}^{2}}p\right) }  \label{ausnahme}
\end{equation}%
solves (\ref{eq}) for the generic Hamiltonian (\ref{hcs}) with the only
constraint that ${{\alpha }_{3}\neq 0}$ and ${{\alpha }_{9}=0}$. In this
case the corresponding Hermitian counterpart is computed to 
\begin{equation}
h_{\text{c}}(x,p)={h}_{0}+g^{2}\frac{{\left( p^{2}\,{{\alpha }_{7}}+p\,{{%
\alpha }_{8}}+{{\alpha }_{10}}\right) }^{2}}{4\,\left( p\,{{\alpha }_{3}}+{{%
\alpha }_{5}}\right) }.  \label{a1}
\end{equation}%
Implementing the constraint (\ref{xy}), the Hamiltonian (\ref{a1}) reduces
to the one in (\ref{a22}). Similarly as the model of the previous
subsection, these solutions can not be reduced to any of the well studied
models presented in table 1, since ${{\alpha }_{3}}$ is assumed to be
non-vanishing.

\subsection{Vanishing $\hat{p}\hat{x}^{2}$-term and non-vanishing $\hat{x}%
^{2}$-term}

Next we consider the complementary case to the previous two section, that is
we take ${{\alpha }_{3}=0}$ in (\ref{Hc}). For this set up we can only find
an exact solution when we demand in addition that ${{\alpha }_{5}\neq 0}$
and ${{\alpha }_{9}=0}$. For the non-Hermitian Hamiltonian 
\begin{equation}
H_{\text{c}}(x,p)=\alpha {_{1}}p^{3}+\alpha {_{2}}p^{2}+\alpha {_{4}}%
p+\alpha {_{5}}x^{2}+\alpha {_{6}}+ig(\alpha {_{7}}p^{2}x+\alpha {_{8}}%
px+\alpha {_{10}}x\,{),}  \label{66}
\end{equation}
we can solve the differential equation (\ref{eq}) exactly by 
\begin{equation}
\eta ^{2}(x,p)=e^{g\,\left( \frac{\,{{\alpha }_{7}}}{3\,{{\alpha }_{5}}}%
p^{3}+\frac{{{\alpha }_{8}}}{2{{\alpha }_{5}}}p^{2}+\frac{{{\alpha }_{10}}}{{%
{\alpha }_{5}}}p\right) }.  \label{cv}
\end{equation}
Once again $\eta ^{2}(x,p)$ only depends on $p$ and we can simply take the
square root to compute $\eta (p)$. Using (\ref{hH}) the corresponding
Hermitian Hamiltonian is subsequently computed to 
\begin{equation}
h_{\text{c}}(x,p)={{\alpha }_{1}}p^{3}+{{\alpha }_{2}}p^{2}+{{\alpha }_{4}}p+%
{{\alpha }_{5}}x^{2}+{{\alpha }_{6}}+g^{2}\frac{{\left( p^{2}{{\alpha }_{7}}%
+p{{\alpha }_{8}}+{{\alpha }_{10}}\right) }^{2}}{4{{\alpha }_{5}}}.
\end{equation}
Obviously these solutions can be reduced to various cases presented in table
1, notably the transformed $-z^{4}$-potential and the Swanson Hamiltonian.

\subsection{Vanishing $\hat{p}\hat{x}^{2}$-term and non-vanishing $\hat{p}$%
-term or non-vanishing $\hat{p}^{2}$-term}

Finally we consider the complementary case of the previous section by taking 
${{\alpha }_{3}=0}$ and allowing ${{\alpha }_{5}}$ to acquire any value. To
be able to find an exact solution we need to impose the additional
constraints 
\begin{equation}
{{\alpha }_{1}={\alpha }_{7}={\alpha }_{9}=0,\qquad }\text{and\qquad }{{%
\alpha }_{4}{\alpha }_{8}=2{\alpha }_{2}{\alpha }_{10},}
\end{equation}%
i.e. we consider the non-Hermitian Hamiltonian 
\begin{equation}
H_{\text{c}}(x,p)=\alpha {_{2}}p^{2}+\alpha {_{4}}p+\alpha {_{5}}%
x^{2}+\alpha {_{6}}+ig\left( {{\alpha }_{8}}px+\alpha {_{10}}x\,\right) {,}
\label{77}
\end{equation}%
for which we can solve equation (\ref{eq}) by 
\begin{eqnarray}
\eta ^{2}(x,p) &=&e^{-g{{\alpha }_{10}/{\alpha }_{4}}x^{2}}\text{ \ \ \ \
for }{{\alpha }_{4}\neq 0,} \\
\eta ^{2}(x,p) &=&e^{-g{{\alpha }_{8}/2{\alpha }_{2}}x^{2}}\text{ \ \ \ \
for }{{\alpha }_{2}\neq 0.}
\end{eqnarray}%
As $\eta ^{2}(x,p)$ only depends on $x,$ we can take the square root to
compute $\eta (x)$ and subsequently evaluate the corresponding Hermitian
counterpart using (\ref{hH}) 
\begin{eqnarray}
h_{\text{c}}(x,p) &=&\alpha {_{2}}p^{2}+\alpha {_{4}}p+\alpha {_{5}}%
x^{2}+\alpha {_{6}}+g^{2}\frac{{{\alpha }_{2}}\alpha {_{10}^{2}}}{\alpha {%
_{4}^{2}}}x^{2}\text{ \ \ \ \ for }{{\alpha }_{4}\neq 0,} \\
h_{\text{c}}(x,p) &=&\alpha {_{2}}p^{2}+\alpha {_{4}}p+\alpha {_{5}}%
x^{2}+\alpha {_{6}}+g^{2}\frac{\alpha {_{8}^{2}}}{4{{\alpha }_{2}}}x^{2}%
\text{ \ \ \ \ for }{{\alpha }_{2}\neq 0.}
\end{eqnarray}%
The Hamiltonian in (\ref{77}) can be reduced to the Swanson Hamiltonian.
Notice that when we impose ${{\alpha }_{1}={\alpha }_{4}={\alpha }_{7}={%
\alpha }_{10}=0}$ for the Hamiltonian in (\ref{66}) and ${{\alpha }_{4}={%
\alpha }_{10}=0}$ for the Hamiltonian in (\ref{77}), they become both
identical to the Swanson Hamiltonian. The corresponding solutions for the
metric operators reduce to $\hat{\eta}^{2}(x,p)=e^{g\,{{\alpha }_{8}/}2{{%
\alpha }_{5}}p^{2}}$ and $\eta ^{2}(x,p)=e^{-g{{\alpha }_{8}/2{\alpha }_{2}}%
x^{2}}$, respectively, which are the well known non-equivalent solutions for
the Swanson Hamiltonian, see e.g. \cite{MGH}. This means according to (\ref%
{symm}) we can identify a symmetry operator for the Swanson Hamiltonian as 
\begin{equation}
S(\hat{x},\hat{p})=e^{-g\,\frac{{{\alpha }_{8}}}{2{{\alpha }_{5}}}\hat{p}%
^{2}}e^{-g\frac{{{\alpha }_{8}}}{{2{\alpha }_{2}}}\hat{x}^{2}}.
\end{equation}%
Notice that $S(x,p)\mathcal{\star }H(x,p)=H(x,p)\mathcal{\star }S(x,p)$ is a
more difficult equation to solve in this example than (\ref{Het}), since $S$
is not of a simple exponential form as $\eta ^{2}$. In fact this is what we
expect. For instance supposing that the symmetry is some group of Lie type,
a typical group element, when Gau$\beta $ decomposed, would be a product of
three exponentials.

\begin{center}
\begin{tabular}{|l||c|c|c|c|c|c|c|c|c|c|}
\hline
model\TEXTsymbol{\backslash}const & $\alpha _{1}$ & $\alpha _{2}$ & $\alpha
_{3}$ & $\alpha _{4}$ & $\alpha _{5}$ & $\alpha _{6}$ & $\alpha _{7}$ & $%
\alpha _{8}$ & $\alpha _{9}$ & $\alpha _{10}$ \\ \hline\hline
$H_{\text{(\ref{HH})}}$ & $\alpha _{1}$ & $\alpha _{2}$ & $\neq 0$ & $\alpha
_{4}$ & $\alpha _{5}$ & $\alpha _{6}$ & $\frac{{{\alpha }_{1}{\alpha }_{9}}}{%
{{\alpha }_{3}}}$ & $\frac{{{\alpha }_{2}{\alpha }_{9}-{\alpha }_{5}{\alpha }%
_{7}}}{{{\alpha }_{3}}}$ & $\alpha _{9}$ & $\frac{{{\alpha }_{4}{\alpha }%
_{9}-{\alpha }_{5}{\alpha }_{8}}}{{{\alpha }_{3}}}$ \\ \hline
$H_{\text{(\ref{H2})}}$ & $0$ & $\alpha _{2}$ & $\neq 0$ & $\alpha _{4}$ & $%
\alpha _{5}$ & $\alpha _{6}$ & $0$ & $\frac{{2{\alpha }_{2}{\alpha }_{9}}}{{{%
\alpha }_{3}}}$ & $\alpha _{9}$ & $\frac{{{\alpha }_{4}{\alpha }_{9}}}{{{%
\alpha }_{3}}}$ \\ \hline
$H_{\text{(\ref{H3})}}$ & $\alpha _{1}$ & $\alpha _{2}$ & $\neq 0$ & $\alpha
_{4}$ & $\alpha _{5}$ & $\alpha _{6}$ & $\alpha _{7}$ & $\alpha _{8}$ & $0$
& $\frac{{{\alpha }_{5}({\alpha }_{3}{\alpha }_{8}-{\alpha }_{5}{\alpha }%
_{7})}}{{{\alpha }_{3}^{2}}}$ \\ \hline
$H_{\text{(\ref{66})}}$ & $\alpha _{1}$ & $\alpha _{2}$ & $0$ & $\alpha _{4}$
& $\neq 0$ & $\alpha _{6}$ & $\alpha _{7}$ & $\alpha _{8}$ & $0$ & $\alpha
_{10}$ \\ \hline
$H_{\text{(\ref{77})}}$ & $0$ & $\alpha _{2}$ & $0$ & $\neq 0$ & $\alpha
_{5} $ & $\alpha _{6}$ & $0$ & $\frac{{2{\alpha }_{2}{\alpha }_{10}}}{\alpha
_{4}} $ & $0$ & $\alpha _{10}$ \\ \hline
$H_{\text{(\ref{77})}}$ & $0$ & $\neq 0$ & $0$ & $\alpha _{4}$ & $\alpha
_{5} $ & $\alpha _{6}$ & $0$ & $\alpha _{8}$ & $0$ & $\frac{\alpha
_{8}\alpha _{4}}{{2{\alpha }_{2}}}$ \\ \hline
\end{tabular}
\end{center}

\noindent {\small Table 2: SPH-models of cubic order.}

Obviously using the relation $\lambda= M^{-1} \alpha$ we can also convert our 
solutions into expressions using creation and annihilation operators.

\section{Lattice version of Reggeon field theory}

Having studied the Hamiltonian (\ref{H211}), (\ref{Hc}) in a very generic
manner let us return to our original motivation and focus on some special
cases, which have hitherto not been dealt with in the literature. It has
been argued for more than thirty years that the lattice versions of Reggeon
field theory (LR) \cite{RR3} 
\begin{equation}
H_{\text{LR}}=\sum\limits_{\vec{\imath}}\left[ \Delta a_{\vec{\imath}%
}^{\dagger }a_{\vec{\imath}}+i\tilde{g}a_{\vec{\imath}}^{\dagger }(a_{\vec{%
\imath}}+a_{\vec{\imath}}^{\dagger })a_{\vec{\imath}}+\hat{g}\sum\limits_{%
\vec{\jmath}}(a_{\vec{\imath}+\vec{\jmath}}^{\dagger }-a_{\vec{\imath}%
}^{\dagger })(a_{\vec{\imath}+\vec{\jmath}}-a_{\vec{\imath}})\right] ~~
\label{Regger}
\end{equation}
with $a_{\vec{\imath}}^{\dagger },a_{\vec{\imath}}$ being standard creation
and annihilation operators and $\Delta ,\hat{g},\tilde{g}\in \mathbb{R}$,
possess a real eigenvalue spectrum \cite{Regger}. This assertion was made
despite the fact that the Hamiltonian in (\ref{Regger}) is non-Hermitian.

\subsection{The single site lattice Reggeon model}

Many features of lattice models can be understood with a finite, possibly
small, number of sites. Thus reducing the Hamiltonian in (\ref{Regger}) to a
single lattice site yields the harmonic oscillator perturbed by a cubic
perturbation in $a$ and $a^{\dagger }$, such that unlike for the so-called
Swanson Hamiltonian its properties can not be understood in direct analogy
to the harmonic oscillator. The resulting model, which we refer to as single
site lattice Reggeon model (SSLR) reads 
\begin{equation}
H_{\text{SSLR}}(a,a^{\dagger })=\Delta a^{\dagger }a+i\tilde{g}a^{\dagger
}\left( a+a^{\dagger }\right) a.  \label{LatReg}
\end{equation}
Obviously $H_{\text{SSLR}}$ is a special case of the generic cubic $\mathcal{%
PT}$-symmetric Hamiltonians (\ref{H211}). One may find already in the old
literature, e.g. \cite{Moshe:1977fe}, that the Hermitian conjugation of $H$
as specified in (\ref{1}) can be obtained by an adjoint action with the
parity operator 
\begin{equation}
\mathcal{P}=e^{\frac{i\pi }{2}a^{\dagger }a}=\eta ^{2}.
\end{equation}
This is easily seen by noting that $\mathcal{P}$ acts on the creation and
annihilation operators as 
\begin{equation}
\mathcal{P}a\mathcal{P}=-a\quad \text{and\quad }\mathcal{P}a^{\dagger }%
\mathcal{P}=-a^{\dagger }.  \label{PP}
\end{equation}
However, since the corresponding $\eta =\sqrt{\mathcal{P}}$ is not a
Hermitian operator, we can not construct a Hermitian counterpart as
specified in (\ref{1}) from this solution for $\eta ^{2}$. Another operator
bilinear in $a$ and $a^{\dagger }$, which has the same effect as $\mathcal{P}
$ in (\ref{PP}) when acting adjointly on $a$,$a^{\dagger }$, is 
\begin{equation}
\hat{\eta}^{2}=e^{\frac{i\pi }{2}(aa-a^{\dagger }a^{\dagger })}.  \label{e2}
\end{equation}
Taking now the square root of (\ref{e2}) yields a Hermitian operator $\hat{%
\eta}$ and serves therefore potentially for our purposes. Noting that this $%
\hat{\eta}$ acts as 
\begin{equation}
\hat{\eta}a\hat{\eta}^{-1}=ia^{\dagger }\quad \text{and\quad }\hat{\eta}%
a^{\dagger }\hat{\eta}^{-1}=ia,  \label{e3}
\end{equation}
such that a corresponding Hermitian counterpart is trivially obtained as 
\begin{equation}
h=\hat{\eta}H_{\text{SSLR}}\hat{\eta}^{-1}=-\Delta aa^{\dagger }+\tilde{g}%
a\left( a+a^{\dagger }\right) a^{\dagger }.
\end{equation}
Clearly the spectrum of this Hamiltonian is not bounded from below for $%
\Delta >0$, resulting essentially from the fact that $\hat{\eta}^{2}$ in not
a legitimate metric operator as it is not positive definite. Nonetheless, in
the context of Reggeon field theory there is a considerable interest in the
regime $\Delta <0$, such that the above argument contributes to an old
discussion.

Ignoring whether the metric is positive-definite or Hermitain we have two
alternative solutions to equation (\ref{1}) involving $H_{\text{SSLR}}$ and
therefore we have simple examples for the symmetry operators $S=$ $\mathcal{P%
}\hat{\eta}^{2}$ and $s=\hat{\eta}\mathcal{P}\hat{\eta}$ in (\ref{symm}) and
(\ref{S2}), respectively.

Let us now try to find a more meaningful metric operator by using Moyal
brackets. To commence, we have to convert the version (\ref{LatReg}) of the
Hamiltonian into one which depends on $\hat{x}$ and $\hat{p}$ instead of $a$
and $a^{\dagger }$. Using the aforementioned relations yields 
\begin{equation}
H_{\text{SSLR}}(\hat{x},\hat{p})=\frac{{\Delta }}{2}(\hat{p}^{2}+\hat{x}%
^{2}-1)+i\frac{\tilde{g}}{\sqrt{2}}{(\hat{x}}^{3}+\hat{p}^{2}\hat{x}-2\hat{x}%
+i\hat{p}{)}\,.  \label{H}
\end{equation}
When ignoring the last three terms, this Hamiltonian becomes the massive
version of the complex cubic potential of Bender and Boettcher \cite%
{Bender:1998ke}. To proceed with our analysis we have to change $H(\hat{x},%
\hat{p})$ from a function depending on operators to a scalar function. In
most terms in (\ref{H}) we can simply replace operators by scalars using $%
\hat{x}\rightarrow x$, $\hat{p}\rightarrow p$, but care needs to be taken
with the term $\hat{p}^{2}\hat{x}\rightarrow p^{2}\star x=p^{2}x-ip$. The
resulting scalar function is 
\begin{equation}
H_{\text{SSLR}}(x,p)=a^{\dagger }\star a+i\tilde{g}a^{\dagger }\star \left(
a+a^{\dagger }\right) \star a=\frac{{1}}{2}(x^{2}+p^{2}-1)+ig{(x}%
^{3}+p^{2}x-2x{),}\,  \label{H1}
\end{equation}
where we have also scaled $\tilde{g}\rightarrow g\sqrt{2}$ and set $\Delta
=1 $. For these values of the constants the differential equation (\ref{eq})
reduces to 
\begin{equation}
\left( 2x\partial _{p}-2p\partial _{x}\right) \eta ^{2}(x,p)=g\left(
4x^{3}-8x+4p^{2}x+2p\partial _{x}\partial _{p}-3x\partial _{p}^{2}-x\partial
_{x}^{2}\right) \eta ^{2}(x,p).  \label{diff}
\end{equation}
Due to its close resemblance to the cubic potential we do not expect this
equation to be exactly solvable and therefore resort to perturbation theory.

\subsection{SPH-models related to the single site lattice Reggeon model}

However, there are various models closely related to $H_{\text{SSLR}}$,
which fit into the scheme of the previous section and are SPH. For instance,
we can identify the exact solution of section 4.1.1. by matching $H_{\text{c}%
}(x,p)$ in (\ref{HH}) with $H_{\text{SSLR}}$ 
\begin{equation}
H_{(\ref{HH})}(a,a^{\dagger },\Delta ,g,\lambda )=H_{\text{SSLR}%
}(a,a^{\dagger },\Delta /2,g)+H_{\text{SSLR}}(-a,a^{\dagger },-\Delta
/2,-\lambda ).  \label{EX1}
\end{equation}%
We find that all the constraints for the ten parameters $\alpha _{i}$ in (%
\ref{con1}) are satisfied for this combination. An exact solution for the
scalar metric function can then be identified as 
\begin{equation}
\eta ^{2}(x,p)=e^{-g/\lambda (x^{2}+p^{2})}.
\end{equation}%
The Hamiltonian in (\ref{EX1}) exhibits an interesting strong-weak symmetry 
\begin{equation}
H_{(\ref{HH})}(a,a^{\dagger },\Delta ,g,\lambda )=H_{(\ref{HH}%
)}(-a,a^{\dagger },-\Delta ,-\lambda ,-g).
\end{equation}%
A further example for a SPH-model related to $H_{\text{SSLR}}$ is 
\begin{equation}
H_{(\ref{66})}(\hat{x},\hat{p},\Delta ,g,\lambda )=H_{\text{SSLR}}(\hat{x},%
\hat{p},\Delta ,g)-i\tilde{g}\hat{x}^{3},  \label{SS3}
\end{equation}%
which is identical to $H_{\text{c}}(x,p)$ in (\ref{66}) for certain values
of the ten parameters $\alpha _{i}$. Reading off those parameters yields as
exact solution 
\begin{equation}
\eta ^{2}(\hat{x},\hat{p})=e^{-\sqrt{2}g/\Delta (\hat{p}^{3}/3-2\hat{p}%
^{2})},
\end{equation}%
for the metric operator according to (\ref{cv}). In fact, this model can be
matched with the transformed version of the $-z^{4}$-potential, for which
the exact metric operator was constructed in \cite{JM}.

\subsection{Perturbative solution}

In order to solve the differential equation (\ref{diff}) perturbatively we
make now the ansatz 
\begin{equation}
\eta ^{2}(x,p)=2\sum\limits_{n=0}^{\infty }g^{n}c_{n}(x,p)
\end{equation}
and the equation (\ref{diff}) is changed into a recursive equation for the
coefficients $c_{n}$ 
\begin{equation}
\left( 2x\partial _{p}-2p\partial _{x}\right) c_{n}(x,p)=\left(
4x^{3}-8x+4p^{2}x+2p\partial _{x}\partial _{p}-3x\partial _{p}^{2}-x\partial
_{x}^{2}\right) c_{n-1}(x,p).  \label{rec}
\end{equation}
We may solve this successively order by order. Using the fact that $%
\lim_{g\rightarrow 0}\eta ^{2}(x,p)=0$ the initial condition is taken to be $%
c_{0}(x,p)=1$. We then obtain recursively order by order 
\begin{eqnarray}
c_{1}(x,p) &=&p^{3}-2p+px^{2}, \\
c_{2}(x,p) &=&p^{6}-4p^{4}+p^{2}+x^{2}-4p^{2}x^{2}+2p^{4}x^{2}+p^{2}x^{4}, 
\notag \\
c_{3}(x,p) &=&\frac{2}{3}%
p^{9}-4p^{7}-5p^{5}+24p^{3}-4p+8px^{2}-6p^{3}x^{2}-8p^{5}x^{2}+2p^{7}x^{2}-px^{4}+%
\frac{2}{3}p^{3}x^{6}  \notag \\
&&-4p^{3}x^{4}+2p^{5}x^{4},  \notag \\
c_{4}(x,p) &=&\frac{1}{3}p^{12}-\frac{8}{3}%
p^{10}-12p^{8}+76p^{6}-5p^{4}-72p^{2}+24x^{2}-18p^{2}x^{2}+104p^{4}x^{2} 
\notag \\
&&-28p^{6}x^{2}-8p^{8}x^{2}-13x^{4}+28p^{2}x^{4}-20p^{4}x^{4}-8p^{6}x^{4}+2p^{8}x^{4}-4p^{2}x^{6}
\notag \\
&&+\frac{4}{3}p^{10}x^{2}-\frac{8}{3}p^{4}x^{6}+\frac{4}{3}p^{6}x^{6}+\frac{1%
}{3}p^{4}x^{8}.  \notag
\end{eqnarray}
Apart from the usual ambiguities, which were discussed in section 3.3, there
are new uncertainties entering through this solution procedure. Apart from
the different choice for the integration constants, it is evident that on
the left hand side of (\ref{rec}) we can always add to $c_{n}(x,p)$ any
function of the Hermitian part of $H_{\text{SSLR}}(x,p)$, i.e. $%
c_{n}(x,p)\rightarrow c_{n}(x,p)+\frac{{1}}{2}(x^{2}+p^{2}-1)$ is also a
solution of the left hand side of (\ref{rec}). We fix this ambiguity by
demanding $\eta ^{2}(x,p,g)\star \eta ^{2}(x,p,-g)=1$ for reasons explained
in \cite{CA}.

Next we solve the differential equation $\eta (x,p)$ $\star \eta (x,p)=\eta
^{2}(x,p)$ for $\eta (x,p)$ by making the ansatz 
\begin{equation}
\eta (x,p)=1+\sum\limits_{n=1}^{\infty }g^{n}q_{n}(x,p).
\end{equation}
We find order by order 
\begin{eqnarray}
q_{1}(x,p) &=&c_{1}(x,p),\qquad q_{2}(x,p)=c_{2}(x,p)/2, \\
q_{3}(x,p) &=&\frac{1}{6}p^{9}-p^{7}-\frac{17}{4}p^{5}+16p^{3}-3p-\frac{15}{2%
}p^{3}x^{2}+\frac{1}{2}p^{7}x^{2}+\frac{p^{5}x^{4}}{2}+\frac{p^{3}x^{6}}{6}-%
\frac{13}{4}px^{4}  \notag \\
&&-p^{3}x^{4}+12px^{2}-2p^{5}x^{2},  \notag \\
q_{4}(x,p) &=&-35p^{2}+\frac{11}{8}p^{4}+\frac{51}{2}p^{6}-\frac{9}{2}p^{8}-%
\frac{1}{3}p^{10}+\frac{1}{24}p^{12}-\frac{25p^{2}x^{2}}{4}+\frac{39}{2}%
p^{2}x^{4}+\frac{1}{24}p^{4}x^{8}  \notag \\
&&+\frac{1}{6}p^{10}x^{2}-\frac{61}{8}x^{4}-\frac{23}{2}p^{4}x^{4}-\frac{25}{%
2}p^{6}x^{2}+\frac{p^{8}x^{4}}{4}-\frac{7}{2}p^{2}x^{6}-\frac{1}{3}%
p^{4}x^{6}+\frac{1}{6}p^{6}x^{6}  \notag \\
&&+45p^{4}x^{2}-p^{8}x^{2}-p^{6}x^{4}+13x^{2}.  \notag
\end{eqnarray}
We are now in the position to compute the Hermitian counterpart to $H_{\text{%
SSLR}}(x,p)$ by means of (\ref{hH}) 
\begin{eqnarray}
h_{\text{SSLR}}(x,p) &=&\frac{{1}}{2}(x^{2}+p^{2}-1)+g^{2}\left( \frac{3}{2}%
p^{4}-4p^{2}+1-4x^{2}+3p^{2}x^{2}+\frac{3}{2}x^{4}\right) \\
&&\!\!\!\!\!\!\!\!\!\!\!\!\!\!\!\!\!\!\!\!\!\!\!\!\!\!\!\!\!\!\!\!\!\!\!\!\!%
\!-g^{4}\left( \frac{17}{2}p^{6}-34p^{4}+4p^{2}+8+4x^{2}-48p^{2}x^{2}+\frac{%
41}{2}p^{4}x^{2}-14x^{4}+\frac{31}{2}p^{2}x^{4}+\frac{7}{2}x^{6}\right) +%
\mathcal{O}(g^{6})  \notag
\end{eqnarray}
Finally we recast our solution again in terms of creation and annihilation
operators. Up to order $g^{2}$ the square root of the metric becomes 
\begin{equation}
\eta =1+i\sqrt{2}ga^{\dagger }(a^{\dagger }-a)a+g^{2}a^{\dagger }\left[
a^{\dagger }(2a^{\dagger }a-a^{\dagger }a^{\dagger }-aa+5)a-2a^{\dagger
}a^{\dagger }-2aa+2\right] a
\end{equation}
and the Hermitian counterpart to the non-Hermitian Hamiltonian $H_{\text{SSLR%
}}$ acquires the form 
\begin{eqnarray}
h_{\text{SSLR}} &=&a^{\dagger }a+g^{2}a^{\dagger }(6a^{\dagger }a+4)a+g^{4}%
\left[ a^{\dagger }a^{\dagger }(10a^{\dagger }a^{\dagger }+10aa-48a^{\dagger
}a)aa\right. \\
&&\text{ \ \ \ \ \ \ \ \ \ \ \ \ \ \ \ \ \ \ \ \ \ \ \ \ \ \ \ \ \ \ \ \ \ \ 
}\left. \text{\ }+a^{\dagger }(20a^{\dagger }a^{\dagger }+20aa-120a^{\dagger
}a)a-32a^{\dagger }a\right] +\mathcal{O}(\hat{g}^{6}).  \notag
\end{eqnarray}
As for all previously constructed perturbative solutions, it would be highly
desirable to investigate in more detail the convergence properties of these
solutions.

\section{Potentials leading to zero cosmological constants and the SSLR-model%
}

For the purpose of identifying vacuum solutions with zero cosmological
constant 't Hooft and Nobbenhuis proposed in \cite{thooft} an interesting
complex space-time symmetry transformation between de-Sitter and
anti-de-Sitter space 
\begin{equation}
\text{dS}\rightarrow \text{adS}:\text{ }x^{\mu }\rightarrow ix^{\mu }~\equiv
~x\rightarrow ix,p\rightarrow -ip~.  \label{ads}
\end{equation}%
Since this transformation relates vacuum solutions with positive
cosmological constant to those with negative cosmological constant, it can
only be a symmetry for the vacuum if the cosmological constant is vanishing.
In order to match this with a quantum mechanical Hamiltonian one demands
that the map dS$\rightarrow $adS sends $H$ to $-H$, such that the vacuum
state is the only invariant state of theory. This means any Hamiltonian of
the form 
\begin{equation}
H_{\text{dS}}(\hat{x},\hat{p})=\sum\nolimits_{j}\hat{x}^{n_{j}}\hat{p}%
^{m_{j}}f_{j}(\hat{x},\hat{p})~~~\text{with }\QATOPD\{ . {n_{j}+m_{j}=4k_{j}%
\text{ \ \ \ for }m_{j}\text{ odd~~~ \ \ \ \ \ \ \ \ \ \ \ }%
}{n_{j}+m_{j}=2k_{j}\text{ \ \ \ for }m_{j}\text{ even, }k_{j}\text{ odd, ~~}%
}  \label{asd}
\end{equation}%
where the $f_{j}(\hat{x},\hat{p})=f_{j}(i\hat{x},-i\hat{p})$ are arbitrary
functions, is respecting this symmetry. A simple example for such a
Hamiltonian was proposed in this context by Jackiw, see \cite{thooft}, 
\begin{equation}
H_{J}(\hat{z},\Omega ,\lambda _{1},\lambda _{2})=\frac{\Omega }{2}\hat{p}%
_{z}^{2}+\lambda _{1}\hat{z}^{6}+\lambda _{2}\hat{z}^{2}.  \label{HJ}
\end{equation}%
Clearly dS$\rightarrow $adS maps $H_{J}\rightarrow -H_{J}$. In fact, his
Hamiltonian involving a sextic potential was investigated before in \cite%
{Bender:1992bk,Bender:1996at,Bender:1998kf}. Notice that $H_{J}$ is also $%
\mathcal{PT}$-symmetric. Furthermore, as pointed out in \cite%
{Bender:1992bk,thooft} for $H_{J}(\hat{z},1,2,-3)$ the groundstate
wavefunction acquires a very simple form $\psi _{0}=\exp (-\hat{z}^{4}/2)$
and $H_{J}$ factorizes, such that it can be interpreted as the bosonic part
of a supersymmetric pair of Hamiltonians. Moreover, this model is
quasi-exactly solvable, meaning that a finite portion of the corresponding
eigensystem has been constructed {Bender:1996at}. As discussed in \cite%
{Bender:1992bk,Bender:1998ke} one can continue the Schr{\"{o}}dinger
equation away from the real axis. Assuming an exponential fall off at
infinity for $H_{J}$ one may choose any parameterization which remains
asymptotically inside the two wedges 
\begin{equation}
\mathcal{W}_{L}=\left\{ \theta \left\vert -\frac{7}{8}\pi <\theta <-\frac{5}{%
8}\pi \right. \right\} \qquad \text{and\qquad }\mathcal{W}_{R}=\left\{
\theta \left\vert -\frac{3}{8}\pi <\theta <-\frac{1}{8}\pi \right. \right\} .
\label{We2}
\end{equation}%
In fact, we can employ the same transformation as the one which was used
successful for the $-z^{4}$-potential in \cite{JM} 
\begin{equation}
~~z(x)=-2i\sqrt{1+ix}.  \label{xz}
\end{equation}%
For large positive $x$ we find $z\sim e^{-i\pi /4}\in \mathcal{W}_{R}$ and
likewise for large negative $x$ we find $z\sim e^{-i\pi 3/4}\in \mathcal{W}%
_{L}$.

\noindent We then find 
\begin{equation}
H_{4}(\hat{x},\hat{p}_{x},g)=H_{4}\left[ \hat{z}(x),\hat{p}_{z},g\right] =%
\frac{\hat{p}_{z(x)}^{2}}{2}-\frac{g}{32}\hat{z}(x)^{4}=\frac{\hat{p}_{x}^{2}%
}{2}+\frac{\hat{p}_{x}}{4}+\frac{g}{2}\hat{x}^{2}-\frac{g}{2}+\frac{i}{2}%
\hat{x}\hat{p}_{x}^{2}-ig\hat{x}.
\end{equation}
This allows us to interpret the Hamiltonian $H_{J}(z,\Omega ,\lambda
_{1},\lambda _{2})$ as a perturbation of the exactly solvable model $H_{4}(%
\hat{z},\hat{p}_{z},g)$, since 
\begin{equation}
H_{J}\left[ \hat{z}(x),1,g/384,g/8\right] =H_{4}(\hat{x},\hat{p}_{x},g)+%
\frac{ig}{6}x^{3}-\frac{g}{6}.
\end{equation}
We can also relate the special case to an exactly solvable model 
\begin{eqnarray}
H_{J}\left[ \hat{z}(x),1,2,-3\right] &=&\frac{\hat{p}_{x}^{2}}{2}-\frac{\hat{%
p}_{x}}{4}+384\hat{x}^{2}-116+i\left( \frac{\left\{ \hat{x},\hat{p}%
_{x}^{2}\right\} }{4}+372\hat{x}+128\hat{x}^{3}\right) \\
&=&H_{(\ref{66})}(\hat{x},\hat{p})+i128\hat{x}^{3}
\end{eqnarray}

It would be very interesting to investigate also the possibility to have a $%
\hat{p}$-dependence in the potential $H_{\text{dS}}(\hat{x},\hat{p})$, but
unfortunately this always leads to equations with order greater than three
and is therefore beyond our generic treatment.

\section{A simple reality proof for the spectrum of $p^{2}+z^{2}(iz)^{2m+1}$}

Considerable efforts have been made to prove the reality of the spectrum for
the family of Hamiltonians $H_{n}=p^{2}+\hat{z}^{2}(i\hat{z})^{n}$ for $%
n\geq 0$, see for instance \cite{DDT,Shin}. Unfortunately most of the proofs
are rather cumbersome and not particularly transparent. The simplest way to
establish the reality of the spectrum for a non-Hermitain Hamiltonian is to
construct exactly the similarity transformation, which relates it to a
Hermitian counterpart in the same similarity class. So far this could only
be achieved for the case $n=2$ in a remarkably simple manner \cite{JM}. Here
we present a trivial argument, which establishes the reality for a subclass
of the massive version of $H_{n}$ with $n$ being odd. Considering 
\begin{eqnarray}
H_{m} &=&\frac{\Delta }{2}(\hat{p}^{2}+\hat{x}^{2})+gx^{2}(i\hat{x}%
)^{2m-1},~~~\ \ \ \ ~~~\Delta ,g\in \mathbb{R},m\in \mathbb{N}, \\
&=&\Delta a^{\dagger }a-\frac{ig(-1)^{m}}{2^{m+1/2}}\left( a+a^{\dagger
}\right) ^{2m+1}  \notag
\end{eqnarray}
it is trivial to see that the metric operator $\hat{\eta}=e^{\frac{i\pi }{4}%
(aa-a^{\dagger }a^{\dagger })}$ introduced in (\ref{e2}) transforms $H_{m}$
adjointly into a Hermitian Hamiltonian 
\begin{align}
h_{m}& =\hat{\eta}H^{m}\hat{\eta}^{-1}=-\Delta aa^{\dagger }+\frac{g}{%
2^{m+1/2}}\left( a+a^{\dagger }\right) ^{2m+1}=h_{m}^{\dagger } \\
& =-\frac{\Delta }{2}(\hat{p}^{2}+\hat{x}^{2})-g\hat{x}^{2m+1}.  \notag
\end{align}
We can immediately apply the same argumentation to generalizations of the
single site lattice Reggeon Hamiltonian. The non-Hermitian Hamiltonian 
\begin{equation}
H_{\text{SSLR}}^{m}=\Delta a^{\dagger }a+\frac{ig}{2}a^{\dagger }\left(
a+a^{\dagger }\right) ^{2m+1}a~~
\end{equation}
is transformed to the Hermitian Hamiltonian 
\begin{equation}
h_{\text{SSLR}}^{m}=\hat{\eta}H_{\text{SSLR}}^{m}\hat{\eta}^{-1}=-\Delta
aa^{\dagger }-\frac{g}{2^{m+1/2}}a^{\dagger }\left( a+a^{\dagger }\right)
^{2m+1}a.
\end{equation}

For the reasons mentioned in section 5 the operator $\hat{\eta}^{2}$ is not
a proper positive-definite metric, but for the purpose of establishing the
reality of the spectrum that is not important.

\section{Conclusions}

We have systematically constructed all exact solutions for the metric
operator which is of exponential form with $\mathcal{PT}$-symmetric real and
cubic argument and adjointly complex conjugates the most generic Hamiltonian
of cubic order (\ref{H211}), (\ref{Hc}). Our solutions are characterized by
various constraints on the ten parameters in the model. Several of the
SPH-models may be reduced to previously studied models, but some correspond
to entirely new examples for SPH-models. We used the metric to construct the
corresponding similarity transformation and its Hermitian counterparts.

We have demonstrated that exploiting the isomorphism between operator and
Moyal products allows to convert the operator identities into manageable
differential equations. Even when no obvious exact solution exists,
perturbation theory can be carried out on the level of the differential
equation to almost any desired order. An important open question, which will
be left for future investigation is concerning the convergence of the
perturbative series. There are obvious limitations for the method as it
works only well for potentials of polynomial form, as otherwise the
differential equations will be of infinite order. In addition even for the
differential equations of finite order not all solutions of all possible
have been obtained as one has to make various assumptions. For instance the
solution (\ref{ausnahme}) would be missed for an ansatz for the metric of a
purely exponential form. Therfore it would be very interesting to compare
the method used in this manuscript with alternative techniques, such as the
one proposed in \cite{MOT}.

The special cases considered, namely the single site lattice Reggeon model (%
\ref{LatReg}) as well as the sextic potential (\ref{HJ}) are both not SPH
within our framework. However, both of them may be understood as quasi
exactly solvable models perturbed by some complex cubic potential.

There are some obvious generalizations of the presented analysis, such as
for instance the study of generic quartic, quintic etc Hamiltonians. \medskip

\noindent \textbf{Acknowledgments}: AF would acknowledge the kind
hospitality granted by the members of the Department of Physics of the
University of Stellenbosch, in particular Hendrik Geyer. AF thanks Carla
Figueira de Morisson Faria for useful discussion. P.E.G.A. is supported by a
City University London research studentship.

\end{document}